\documentclass[preprint]{aastex}
\shorttitle{LiH depletion}
\shortauthors{Bovino et al.}
\begin{document}
\title{On the relative abundance of LiH and LiH$^+$ molecules in the early universe: 
new results from quantum reactions}
\author{Stefano~Bovino, Mario~Tacconi, Franco~A.~Gianturco}
\affil{Department of Chemistry and CNISM, the University of Rome ``Sapienza'', \\ 
P.le A. Moro 5, 00185 Roma, Italy}
\author{Daniele~Galli, Francesco~Palla}
\affil{INAF-Osservatorio Astrofisico di Arcetri,\\Largo E. Fermi 5, 50125 Firenze, Italy}

\email{fa.gianturco@caspur.it}

\begin{abstract}
The relative efficiencies of the chemical pathways that can lead to the
destruction of LiH and LiH$^+$ molecules, conjectured to be present in
the primordial gas and to control molecular cooling processes in the
gravitational collapse of the post-recombination era, are revisited by
using accurate quantum calculations for the several reactions involved.
The new rates are employed to survey the behavior of the relative
abundance of these molecules at redshifts of interest for early
universe conditions.  We find significant differences with respect to
previous calculations, the present ones yielding LIH abundances higher
than LiH$^+$ at all redshifts.
\end{abstract}

\section{Introduction}

The suggestion that chemical processes involving lithium could play a
role in the evolution of the early universe in the post-recombination
era has been put forward several years ago (Lepp \& Shull~1983; Stancil
et al. 1996, hereafter SLD96; Galli \& Palla 1998; hereafter GP98),
starting from the consideration that Li is produced a few minutes after
the Big Bang and the formation of other light atoms like H, D and He
(Wagoner et al. 1967, Peebles 1993). The fractional abundances of these
elements are sensitive to the baryon density of the universe and set
constraints on its actual value (Cyburt et al. 2008). As the universe
expanded, its radiation temperature decreased and the atomic ions
originating from the above elements gave rise to neutral atoms by
recombination with electrons, thus initiating the formation of
molecular species by radiative association: H$_2$, HD and LiH. The
latter molecule, because of its large dipole moment and low ionization
potential, may induce spatial and/or spectral distortions in the cosmic
background radiation (CBR), as originally suggested by
Dubrovich~(1993) and as experimentally surmised by the pionering
work of Maoli et al. (1994). Eventually, during the gravitational collapse leading
to the formation of the first stars, the low excitation threshold and
the efficient radiative decay along the rotovibrational manifold are
additional properties that could favor the role of LiH and LiH$^+$ as
molecular coolants of the primordial gas (Bougleux \& Galli 1997,
hereafter BG97).  The nonequilibrium level population in the presence
of a different gas and radiation temperature may have a possible
signature in protogalactic clouds, imprinting small fluctuations in the
CBR spectrum (Schleicher et al. 2008), as we shall further discuss in our
conclusions.

The chemistry of Li in the early universe has been discussed in
the past, reaching contrasting conclusions (see e.g., SLD96; BG97; GP98;
Vonlanthen et al. 2009). Of critical relevance, is the uncertainty in
the knowledge of reliable reaction rates for the destruction of LiH and
LiH$^+$ molecules 
via strongly exothermic reactions
without entrance barriers:
\begin{equation}
\rm LiH + H \rightarrow Li + H_2,
\end{equation}
\begin{equation}
\rm LiH + H^+ \rightarrow Li + H_2^+,
\end{equation}
and 
\begin{equation}
\rm LiH^+ + H \rightarrow Li^+ + H_2.
\end{equation}
Therefore, it is an accurate knowledge of the reaction rates for the
above processes, at low redshift values, that can ultimately tell
us what the end-role of the LiH/LiH$^+$ systems could be as efficient
coolants under early universe conditions. The task of the present work is
to show that the reaction rates recently determined from fully ab-initio
quantum methods (Bovino et al.~2009, 2010a, 2010b), which also employ
accurate interaction forces between partners, have a significant impact
on the evolution of LiH and LiH$^+$ during the post-recombination era of
the early universe. We shall further show that a more realistic computational description
of the rates for a neutralization process
\begin{equation}
\rm LiH^+ + e^- \rightarrow Li + H
\end{equation}
could substantially change the relative abundance of the ionic molecular species.

This paper is organized as follows: the details of the quantum methods
for the new chemical reaction rates are given in Section~2, while
Section~3 reports the list of considered reactions and describes the
equations controlling the evolution of the gas temperature. In Section~4
we present the fractional abundances and an analysis of their
behavior, while Section~5 summarizes our present conclusions.

\section[]{The quantum reactive calculations}

The study of the chemical
evolution of the main species of the primordial gas requires accurate
rate coefficients for the most important destruction reactions. In
previous works, the Langevin approximation for barrierless reactions
has been the main method used to obtain the rate coefficients for those
reactions. In the last decade, with the developing of more realistic
methods for reaction processes, a great deal of work has been done to
study reactions (1), (2) and (3). For instance, time-dependent (Bulut
et al. 2008, Defazio et al. 2005) and quasi-classical trajectory (Pino
et al. 2008, Dunne et al. 2001) calculations have been reported, and
some of the calculated rates have been included in the abundance
calculations of Lepp et. al.~(2002). However, most of these
calculations are restricted to a narrow range of energies or are
limited to the high energy regime.  Therefore, the main improvement of
the present calculations has been the use of new, very accurate,
potential energy surfaces (Martinazzo et al. 2003a, 2003b, Wernli et
al. 2009) combined with accurate quantum reactive calculations. The
quantum methods range from the fully ab-initio coupled channels
(Bovino et al.~2009) to the use of a negative imaginary potential 
approach (Bovino et al.~2010a, 2010b) for the ionic systems. The
computed reaction probabilities and cross sections have been reported
in previous papers and we shall not repeat here the details of the
calculations. The final state-to-all cross section obtained from the
probability ($P^{J}_{a\rightarrow all}$) is expressed as

\begin{equation}
\sigma_{(a\rightarrow all)}(E) = 
\frac{\pi}{(2j_a + 1)k_a^2}\sum_J(2J + 1) P^{J}_{a\rightarrow all},
\end{equation}
where $E$ is the collisional energy, $j_a$ the molecular rotational angular momentum, 
$J$ the total angular momentum, and $k_a^2$ is the wave vector defined as
\begin{equation}
k_a^2 = \frac{2\mu}{\hbar^2}(E - \varepsilon_a),
\end{equation}
and $\varepsilon_a$ is the $a$-channel energy.
Rate coefficients are computed by averaging the appropriate reactive
cross sections over a Boltzmann distribution of velocities for the
incoming atom:
\begin{equation}
k(T_{\rm g}) = \frac{(8k_BT_{\rm g}/\pi\mu)^{1/2}}{(k_BT_{\rm g})^2}
\int_0^\infty\sigma_{a\rightarrow all}(E)\exp(-E/k_BT_{\rm g})E\,dE
\end{equation}
where $T_{\rm g}$ is the gas temperature, $k_B$ is the Boltzmann constant and
$\mu$ the reduced mass of the system. The newly computed rate coefficients
for reactions (1), (2) and (3) are reported in Bovino et al.~(2009,
2010a, 2010b) and shown in Figure~1.
The corresponding temperature-dependent fits are given in Table~1, along with all the reactions 
rates for the Li species employed in the present work.

\section{The chemical network and the evolutionary modelling}

The evolution of the pregalactic gas is considered in the framework of
a Friedmann cosmological model and the primordial abundances of the
main gas components are taken from the standard big bang
nucleosynthesis results (Smith et al. 1993). The numerical values of
the cosmological parameters used in the calculation are obtained from
WMAP5 data (Komatsu et al.~2009) and are listed in Table~2.


In order to calculate the abundances of LiH and LiH$^+$,
a set of differential coupled chemical rate equations of the form:
\begin{equation}
\frac{dn_i}{dt} = k_{\rm form}n_jn_k -k_{\rm dest}n_i + \ldots
\label{evol}
\end{equation}
has been solved. In eq.~(8), $k_{\rm form}$ and $k_{\rm dest}$
are the formation and destruction reaction rates as listed in Table~1,
and $n_i$ is the number density of the reactant species $i$. The evolution
of the gas temperature $T_{\rm g}$ is governed by the equation (see e.g. GP98):
\begin{equation}
\frac{dT_{\rm g}}{dt} = -2T_{\rm g}\frac{\dot{R}}{R} 
+ \frac{2}{3kn}[(\Gamma - \Lambda)_{\rm Compton} + (\Gamma - \Lambda)_{\rm mol}],
\end{equation}
where the first term represents the adiabatic cooling associated with
the expansion of the Universe, $R$ being the scale factor. The other two
terms represent, respectively, the net transfer of energy from the CBR to
the gas (per unit time and unit volume) via Compton scattering of CBR
photons and electrons,
\begin{equation}
(\Gamma - \Lambda)_{\rm Compton} = n_e\frac{4k\sigma_TaT_{\rm r}^4(T_{\rm r} - T_{\rm g})}{m_ec},
\end{equation}
and via excitation and de-excitation of molecular transition
\begin{equation}
(\Gamma - \Lambda)_{\rm mol} = \sum_kn_k\sum_{i>j}(x_iC_{ij} - x_jC_{ji})h\nu_{ij},
\end{equation}
where $C_{ij}$ and $C_{ji}$ are the collisional excitation and
de-excitation coefficients and $x_i$ are the fractional level
populations: for more details of this model see BG97 and GP98.
The radiation temperature is $T_{\rm r}(z)=T_0(1+z)$, where $T_0$ is
the present-day CBR temperature. The chemical/thermal network is then completed
by the rate equation for the redshift
\begin{equation}
\frac{dt}{dz} = -\frac{1}{(1 + z) H(z)},
\end{equation}
where
\begin{equation}
H(z)=H_0\sqrt{\Omega_{\rm r}(1+z)^4+\Omega_{\rm m}(1+z)^3+\Omega_{\rm K}(1+z)^2+\Omega_\Lambda}.
\end{equation}
Here, $H_0$ is the Hubble constant and $\Omega_{\rm r},\Omega_{\rm m},\Omega_{\rm K},\Omega_\Lambda$ 
are density parameters. The density $n(z)$ of baryons at redshift $z$ is 
\begin{equation}
n(z) = \Omega_{\rm b} n_{\rm cr}(1 + z)^3,
\end{equation}
with $n_{\rm cr}=3H_0^2/(8\pi G m_{\rm H})$ being the critical
density.  The calculations have been carried out from $z = 10^4$ down
to $z = 1$ and are discussed in the next Section.


\section{Results and discussion}

As discussed in the Introduction, one of the novelties of the present
work is the use, for some of the most important chemical processes, of
rates obtained from accurate quantum calculations and not from
qualitative estimates.  It is therefore important for the reactions
involving LiH/LiH$^+$ to recall the behavior of the ab-initio reactive
rates (in the temperature range of interest) that
have been obtained via the quantum methods outlined in Section~2.
The data reported in Figure~1 indicate the depletion rates for the
disappearance of LiH molecules by exothermically reacting with the
surrounding atomic hydrogen (top panel). The temperature behavior of
the ab-initio data is clearly very different from that assumed by
SLD96.  The new calculations of Bovino et al.~(2009) indicate a
variation over the same range of nearly three orders of magnitude.  The
same occurs for the destruction reaction of LiH by interaction with
protons (middle panel):  the ab-initio calculations (Bovino et al.
2010b) show that the rate varies by about one order of magnitude over
the same range of temperatures.  Finally, the quantum rates associated
with the destruction of LiH$^+$ by the surrounding hydrogen gas (bottom
panel) confirm the weak temperature dependence reported earlier (Lepp
et al.~2002), although they turn out to be about a factor of 4 larger
than the estimate of SLD96.  Once the new quantities are employed
within the evolutionary scheme outlined in Section~3, they yield
a new set of production/destruction rates of LiH and LiH$^+$ as a
function of redshift, as shown in Figure~2 and 3.

\subsection{LiH chemistry}

The various curves shown in Figure~2 follow the notation given in
Table~1 to the most relevant processes: those labelled as $p_1, \ldots,
p_4$ (solid lines) refer to the different reactions leading to LiH
production, while those labelled $d_1, \ldots, d_4$ (dashed lines)
refer to the destruction processes.  A look at the figure allows one to
make the following comments:

({\it i}\/) The production by radiative recombination from the Li($^2P$)
state ($p_1$) clearly dominates at high redshifts, while the production by
recombination from the Li($^2S$) state ($p_2$) becomes the dominant route from
$z\simeq 800$ down to $z\simeq 80$. Then, the channel of recombination
driven by electron attachment ($p_4$) takes over.  However, we must note that the
latter rate is not known from quantum calculations, but only estimated
by SLD96, together with the estimates of LiH formation from H$^-$
(SLD96).

({\it ii}\/) At the highest redshifts, the destruction channel is
dominated by photodissociation ($d_1$), a rate estimated from the
detailed balance of process ($p_2$), while at lower $z$ values
($z\lesssim 300$) the collisional process of reactive destruction
discussed earlier ($d_2$) takes over and dominates the whole
destruction channel down to the lowest redshifts.  The same
reactive process with protons ($d_3$) is seen to be less important,
although larger than previous estimates ($d_4$).

In conclusion, the above data indicate that the radiative
photoassociation from the $n=25$ level of Li is the most significant
formation path, while the collisional destruction by H ($d_2$) 
overcomes production over the whole range of redshifts below 300.

\subsection{LiH$^+$ chemistry}

Figure~3 shows the processes leading to the formation and the
destruction of LiH$^+$, labelled as $p_1^+, \ldots, p_2^+$ and $d_1^+,
\ldots, d_4^+$ as in Table~1. As clearly shown by the figure, the
radiative association processes ($p_1^+$ and $p_2^+$) are seen to
dominate the formation of LiH$^+$ over a very broad range of redshifts.
The association involving the Li ion ($p_1^+$) is the major pathway
down to about $z\simeq 700$, when the reaction of H$^+$ with neutral Li
($p_2^+$) takes over and becomes dominant by several orders of
magnitude over the rates of the charge-exchange reaction ($p_3^+$)
estimated by SLD96.

As for the destruction processes for LiH$^+$, one sees that
photodissociation ($d_1^+$) is the dominant destruction path of the
ionic molecule, due to its small dissociation energy value.
Furthermore, the dominant dissociation channel comes from Li$^+$
production ($d^+_3$), while the one yielding H$^+$ ($d_2^+$) is
relevant only at the highest redshifts. The electron-induced
dissociation ($d_4^+$) was earlier based on a simple estimates from
SLD96 while now we have included the more accurate quantum calculations
from \v{C}urik \& Greene (2007,2008) which yield much larger rates.  It
turns out to be important only at low redshifts.  On the other
hand, the chemical paths which involve the reactions that destroy
LiH$^+$ ($d_3^+$) are seen to be indeed the most important processes
around $z\simeq 40$ while becoming less efficient than the $d^+_4$
process below $z\simeq 10$.  The electron-assisted dissociation of
LiH$^+$ ($d_4^+$) is therefore important only at the lowest redshifts.

In summary, the ionic formation is dominated by the radiative
association of Li with H$^+$ over a very broad range of redshifts,
while destruction by photodissociation prevails at larger redshift and
chemical destruction ($d_3^+$) dominates at lower redshifts.  However,
down to $z\simeq 1$ the dissociative neutralization ($d^+_4$) is
markedly dominant.

\subsection{Chemical evolution}

The new evolution of relative abundances of atomic and molecular
species involving Li is displayed in Figure 4, comparing the earlier
results obtained with the chemical network of GP98 (dashed lines) and
the present results (solid lines) which employ the new quantum rates.
The main results are the following:

({\it i}\/) The temperature dependence of the quantum rates, shown by
the data of Figure~1, is clearly reflected in the molecular abundances
of LiH at redshifts below $z\sim 300$. The rapid rise of LiH at
$z \sim 300$ found by GP98 is no longer seen owing to the increased
destruction rate of channel (1). The rise at $z\sim 100$ is due to the
importance of LiH formation by Li$^-$ that persists at all redshifts
together with the radiative association of Li with H. This is
drastically different from the behavior found by GP98 where the
constant value of the rate for reaction (1) caused the steady drop of
LiH.  The final ($z=0$) value of LiH is $\sim 7\times 10^{-18}$ with an
increase of a factor of $\sim 70$ relative to GP98.


({\it ii}\/) Due to the differences in behavior shown by the cross
sections in Figure~1, the ionic partner, LiH$^+$, behaves differently
at $z\lesssim 30$. First, the increased efficiency of LiH$^+$
destruction (reaction 3; rate $d_3^+$) limits the sharp rise in
abundance at $z \sim 20$. Then, the gentle decline below $z \sim 5$ is due
to the efficiency of electronic recombination (rate $d_4^+$).  The new
final abundances of LiH$^+$ (solid line) is now smaller than earlier
estimates of GP98 by a factor of $\sim 20$.

The scenario emerging from the above calculations therefore indicates
that in the regions of redshift below $z\simeq 30$, LiH remains the
more abundant species compared to LiH$^+$, but only by a factor of
$\sim 2$--7. The two molecular abundances, on the other hand, reach the
largest values for $z\lesssim 10$, remaining both fairly small
(LiH$\sim$10$^{-17}$, LiH$^+ \sim$10$^{-18}$) and hard to detect.


\section{Conclusions}

Building upon recent quantum reactive calculations (Bovino et al. 2009,
2010a, 2010b) involving the chemical evolution of LiH/LiH$^+$ molecules
within the expected conditions in the early universe, the present work
has revisited the analysis of all the dynamical processes that are
known to significantly contribute to the production/destruction of the
lithium-containing molecules. We have employed as many results as
possible from calculations based on ab-initio methods, both for the
interaction forces and the quantum dynamics, resorting to estimates
only for a few of the considered processes.

One of the main results from the quantum reactive calculations is that
two of the important rates exhibit a temperature dependence that was
not present in the earlier estimates.

The results for the abundance of LiH indicate that this molecule is now
much more likely to have survived at low redshift, since its fractional
abundance, albeit still fairly small, goes up by a factor of $\sim 70$
compared to previous estimates. However, it becomes smaller by nearly
the same amount for $z>300$.

The fractional abundance of LiH$^+$ only becomes significant in the
low-redshift region of $z \lesssim 30$ and shows a reduced value of the
relative abundance by about one order of magnitude with respect to the
earlier estimates. The comparison between the specific fractional
abundances of the two species now indicates that the neutral molecule
is likely to be more abundant than the ionic species and that their
ratio in the region of small redshift increases up to a factor of 7
down to $z\simeq 1$. Unlike previous estimates that predicted a
difference at low-$z$ of about two orders of magnitude in favor of
LiH$^+$, the new calculations find more comparable abundances of LiH
and LiH$^+$. Furthermore, we find that the neutral molecule, in spite
of a large dilution at the low redshifts, should be more amenable to
experimental observation (e.g., Persson et al. 2010) than its ionic
counterpart.

Finally, given the recent studies (e.g. see Zaldagarriaga and Loeb,
2002) on the possible detection, wihin the microwave background
anisotropies, of the earlier imprint from the recombination history of
lithium, the present results strongly suggest that such anisotropies
could be amenable to observation due to the changed optical depth
induced by the changes on lithium abundances

\acknowledgments
{\bf Acknowledgments}
We thank the CINECA and CASPUR consortia for providing us with the
necessary computational facilities and the University of Roma
``Sapienza'' for partial financial support.

\begin{center}
{\bf REFERENCES}
\end{center}

\noindent Bennett O.J., Dickinson, A.S., Leininger, T., Gad\'ea, F.X. 2003, MNRAS, 341, 361\\
Bennett O. J., Dickinson A. S., Leininger T. \& Gad\'ea F. X., 2008, MNRAS, 384, 1743\\
Bougleux E. \& Galli D., 1997, MNRAS., 288, 638 (BG97)\\
Bovino S., Wernli M. \& Gianturco F. A., 2009, ApJ, 699, 383\\
Bovino S., Stoecklin T. \& Gianturco F. A., 2010a, ApJ, 708, 1560\\
Bovino S., Tacconi M., Gianturco F. A. \& Stoecklin T., 2010b, ApJ, 724, 106\\
Bulut N., Castillo J. F., Aoiz F. J., \& Banares L., 2008, Phys. Chem. Chem. Phys., 10, 821\\
Cyburt R. M., Fields B. D. \& Olive K. A., 2008, J. Cosmol. Astropart. Phys., 11, 12\\
Croft H., Dickinson A. S. \& Gad\'ea F. X., 1999, MNRAS, 304, 327\\
Cur\'ik R. \& Greene C. H., 2007, Phys. Rev. Lett., 98, 173201\\
Cur\'ik R. \& Greene C. H., 2008, J. Phys.: Conference Series, 115, 012016\\
Dalgarno A., Kirby K. \& Stancil P. C., 1996, ApJ, 458, 397\\
Defazio P., Petrongolo C., Gamallo P. \& Gonz\'alez M., 2005, J. Chem. Phys., 122, 214303\\
Dubrovich V. K., 1993, Astron. Lett., 19, 53\\
Dunne L. J., Murrell J. N. \& Jemmer P.,2001, Chem. Phys. Lett., 336, 1\\
Galli D., Palla F., 1998, A\&A, 335, 403 (GP98)\\
Gianturco F. A. \& Gori Giorgi P., 1996, Phys. Rev. A., 54, 1\\
Gianturco F. A. \& Gori Giorgi P., 1997, ApJ, 479, 560\\
Kimura M., Dutta C. M. \& Shimakura N., 1994, ApJ, 430, 435\\
Komatsu, E., et al. 2009, ApJS, 180, 330 \\
Lepp, S. \& Shull, J.M. 1983, ApJ, 270, 578\\
Lepp S., Stancil P. C. \& Dalgarno A., 2002, J. Phys. B:At. Mol. Opt. Phys., 35, R57-R80\\
Maoli R., Melchiorri F., and Tosti D., 1994, \apj, 425, 372\\
Martinazzo R., Bodo E., Gianturco F. A., \& Raimondi M., 2003a, Chem. Phys., 287, 335\\
Martinazzo R., Tantardini G. F., Bodo E., \& Gianturco F. A., 2003b, J. Chem. Phys., 119, 11241\\
Peebles P. J. E., 1993, Principles of Physical Cosmology, Princeton University Press, Princeton\\
Persson, C.~M., et al.\ 2010, \aap, 515, A72\\
Pino I., Martinazzo R., \& Tantardini G., 2008, Phys. Chem. Chem. Phys., 10, 5545\\
Ramsbottom C. A., Bell K. L. \& Berrington K. A., 1994, J. Phys. B, 27, 2905\\
Smith M. S., Kawano L. H. \& Malaney R. A., 1993, ApJS, 85, 219\\
Schleicher, D.~R.~G., Galli, D., Palla, F., Camenzind, M., Klessen, R.~S., Bartelmann, M., \& Glover, S.~C.~O.\ 2008, \aap, 490, 521\\
Spergel D.~N., et al. 2007, ApJS, 170, 377 \\
Stancil P. C., Lepp S. \& Dalgarno A., 1996, ApJ, 458, 401 (SLD96)\\
Stancil P. C. \& Zygelman B., 1996, ApJ, 472, 102\\
Verner D. A., Ferland G. J., 1996, ApJS, 103, 467\\
Vonlanthen, P., Rauscher, T., Winteler, C., Puy, D., Signore, M., Dubrovich, V. 2009, A\&A, 503, 47\\
Wagoner R. V., Fowler W. A. \& Hoyle F., 1967, ApJ, 148, 3\\
Wernli M., Caruso D., Bodo E. \& Gianturco F. A., 2009, J. Phys. Chem. A., 113, 1121\\
Zaldagarriaga M., \& Loeb A., 2002, \apj, 564, 52\\

\begin{table*}
\tiny
\caption{List of considered reactions. Labels in the second column refer to 
Figures~2 and 3.}
\begin{flushleft}
\begin{tabular}{llllll}
\hline
& &  reaction   &   rate (cm$^3$~s$^{-1}$ or s$^{-1}$)   & notes &  reference \\
\hline
& & & & & \\
\multicolumn{6}{c}{Li reactions} \\

1) & $p_1$ & Li$(^2p)$ + H $\rightarrow$ LiH + $h\nu$ &
      $2.0\times 10^{-16}T_{\rm g}^{0.18}\exp(-T_{\rm g}/5100)$ &
      $A^1\Sigma^+\rightarrow X^1\Sigma^+$, quantal calc. & (a) \\
2) & $p_1$ & Li$(^2p)$ + H $\rightarrow$ LiH + $h\nu$ &
      $1.9\times 10^{-14}T_{\rm g}^{-0.34}$ &
      $B^1\Pi\rightarrow X^1\Sigma^+$, quantal calc. & (a) \\
3) & $p_2$ & Li + H $\rightarrow$ LiH + $h\nu$ &
      $4.0\times 10^{-20}\times$ & & \\
& & & $\exp[-T_{\rm g}/4065+(T_{\rm g}/13193)^3]$
   & quantal calc. & (b),(c),(d) \\
4) & $p_3$ & Li + H$^-$ $\rightarrow$ LiH + e                 &
      $4.0\times 10^{-10}$ & estimate & (e) \\
5) & $p_2^+$ & Li + H$^+$ $\rightarrow$ LiH$^+$ + $h\nu$ &
      $5.3\times 10^{-14} T_{\rm g}^{-0.49}$ 
      & quantal calc. & (c),(d) \\
6) & & Li + H$^+$ $\rightarrow$ Li$^+$ + H               &
      $2.5\times 10^{-40}T_{\rm g}^{7.9}\exp(-T_{\rm g}/1210)$
    & quantal calc. & (f) \\
7) & & Li + H$^+$ $\rightarrow$ Li$^+$ + H + $h\nu$      &
      $1.7\times 10^{-13}T_{\rm g}^{-0.051}\exp(-T_{\rm g}/282000)$
      & quantal calc. & (g) \\
8) & & Li + e $\rightarrow$ Li$^-$ + $h\nu$   &
      $6.1\times 10^{-17}T_{\rm g}^{0.58}\exp(-T_{\rm g}/17200)$
      & det. bal. applied to (27) &  \\
9) & & Li + H$_2^+$ $\rightarrow$ LiH + H$^+$  &
      $6.3\times 10^{-10} \exp(-2553/T_{\rm g})$ 
      & $T_{\rm g} < 500$, quantal calc. & (i) \\
&   &   & 
      $7.2\times 10^{-14} T_{\rm g}^{1.18}\exp(-1470/T_{\rm g})$
      & $T_{\rm g} > 500$, quantal calc. & (j) \\

& & & & & \\
\multicolumn{6}{c}{Li$^+$ reactions} \\

10) & $p_1^+$  & Li$^+$ + H $\rightarrow$ LiH$^+$ + $h\nu$         &
      $1.4\times 10^{-20}T_{\rm g}^{-0.9}\exp(-T_{\rm g}/7000)$
    & quantal calc. & (c),(d) \\
11) & & Li$^+$ + e $\rightarrow$ Li + $h\nu$           &
       $1.036\times 10^{-11}[\sqrt{T_{\rm g}/107.7}\times$ & & \\
   & & & $(1+\sqrt{T_{\rm g}/107.7})^{0.6612}\times$ & & \\
   & & & $(1+\sqrt{T_{\rm g}/1.177\times 10^7})^{1.3388}]^{-1}$
   & quantal calc. & (k) \\
12) &  & Li$^+$ + H$^-$ $\rightarrow$ Li + H &
      $6.3\times 10^{-9}T_{\rm g}^{-1/2}(1+T_{\rm g}/14000)$
      & Landau-Zener approx. &  (l) \\

& & & & & \\
\multicolumn{6}{c}{Li$^-$ reactions} \\

13) & $p_4$ & Li$^-$ + H $\rightarrow$ LiH + e                 &
      $4.0\times 10^{-10}$ & estimate & (e) \\
14) & & Li$^-$ + H$^+$ $\rightarrow$ Li + H         & 
      $2.3\times 10^{-6} T_{\rm g}^{-1/2}$      
      & Landau-Zener approx. & (l) \\

& & & & & \\
\multicolumn{6}{c}{LiH reactions} \\

15) & $d_2$ & LiH + H $\rightarrow$ Li + H$_2$                 &
      $2.0\times 10^{-12} T_{\rm g} \exp(-T_{\rm g}/1200)$ & 
      quantal calc. & (m) \\
16) & $d_3$ & LiH + H$^+$ $\rightarrow$ Li + H$_2^+$     &
      $2.9\times 10^{-10} T_{\rm g}^{0.59}$
      &  & \\ 
& & & $-2.6\times 10^{-10} T_{\rm g}^{0.60}\exp(-400/T_{\rm g})$
      & quantal calc. & (n) \\
17) & $p_3^+$, $d_4$ & LiH + H$^+$ $\rightarrow$ LiH$^+$ + H            &
     $1.0\times 10^{-9}$ & quantal calc. & (o) \\
18) & $d_4$ & LiH + H$^+$ $\rightarrow$ Li$^+$ + H$_2$         &
      $1.0\times 10^{-9}$ & estimate & (e) \\

& & & & & \\
\multicolumn{6}{c}{LiH$^+$ reactions} \\

19) & $d_3^+$ & LiH$^+$ + H $\rightarrow$ Li$^+$ + H$_2$         &
      $8.7\times 10^{-10}T_{\rm g}^{0.040}\exp(T_{\rm g}/5.92\times 10^8)$ 
      & quantal calc. & (p) \\
20) & $d_3^+$ & LiH$^+$ + H $\rightarrow$ Li + H$_2^+$           &
      $9.0\times 10^{-10}\exp(-66400/T_{\rm g})$ & estimate & (e) \\
21) & $d_3^+$ & LiH$^+$ + H $\rightarrow$ LiH + H$^+$            &
      $1.0\times 10^{-11}\exp(-67900/T_{\rm g})$ & estimate & (e) \\
22) & $d_4^+$ & LiH$^+$ + e $\rightarrow$ Li + H                 &
      $3.9\times 10^{-6}T_{\rm g}^{-0.70}\exp(-T_{\rm g}/1200)$ & quantal calc.  & (q) \\

& & & & & \\
\multicolumn{6}{c}{photoreactions} \\

23) & $d_1$ & LiH + $h\nu$ $\rightarrow$ Li + H   &
      & det. bal. applied to (3) &  \\
24) & $d_1^+$ & LiH$^+$ + $h\nu$ $\rightarrow$ Li$^+$ + H &
    & det. bal. applied to (10) &  \\
25) & $d_2^+$ & LiH$^+$ + $h\nu$ $\rightarrow$ Li + H$^+$ &
      & det. bal. applied to (5) &  \\
26) & & Li + $h\nu$ $\rightarrow$ Li$^+$ + e           &
      & det. bal. applied to (11) &  \\
27) & & Li$^-$ + $h\nu$ $\rightarrow$ Li + e  &
      & quantal calc.&  (h) \\
28) & & LiH + $h\nu$ $\rightarrow$ Li$(^2p)$ + H &
      & det. bal. applied to (1), (2) &  \\
& & & & & \\
\hline
\end{tabular}

\vspace{1em}

(a)~Gianturco \& Gori Giorgi~(1996); (b)~Bennett et al.~(2003,
2008); (c)~Dalgarno, Kirby \& Stancil~(1996); (d)~Gianturco \& Gori
Giorgi~(1997); (e)~Stancil, Lepp \& Dalgarno~(1996); (f)~Kimura, Dutta
\& Shimakura~(1994); (g)~Stancil \& Zygelman~(1996); (h)~Ramsbottom,
Bell \& Berrington~(1994); (i)~Bulut et al.~(2009); (j)~Pino et
al.~(2008); (k)~Verner \& Ferland~(1996); (l)~Croft, Dickinson \&
Gad\'ea~(1999); (m)~Bovino, Wernli \& Gianturco~(2009); (n)~Bovino et
al.~(2010); (o)~Bulut et al.~(2008); (p)~Bovino, Stoecklin \& Gianturco~(2010); (q)~\v{C}urik \&
Greene~(2007, 2008). 
\end{flushleft}
\end{table*}

\begin{table*}
\begin{flushleft}
\caption{Cosmological model}
\begin{tabular}{ll}
\hline
Parameter & Numerical value\\
\hline 
$H_0$ & $70.5$~km~s$^{-1}$~Mpc$^{-1}$ \\
$T_0$ & $2.725$~K\\
$z_{\rm eq}$ & $3141$\\
$\Omega_{\rm dm}$ & $0.228$\\
$\Omega_{\rm b}$ & $0.0456$\\
$\Omega_{\rm m}$ & $\Omega_{\rm dm}+\Omega_{\rm b}$\\
$\Omega_{\rm r}$ & $\Omega_{m}/(1+z_{\rm eq})$\\
$\Omega_\Lambda$ & $0.726$\\
$\Omega_{\rm K}$ & $1-\Omega_{\rm r}-\Omega_{\rm m}-\Omega_\Lambda$ \\
$f_{\rm H}$ &  0.924\\
$f_{\rm He}$ & 0.076\\
$f_{\rm D}$ &  $2.38\times 10^{-5}$\\ 
$f_{\rm Li}$ &  $4.04\times 10^{-10}$\\ 
\hline
\end{tabular}
\vspace{1em}
\end{flushleft}
\end{table*}

\begin{center}
\begin{figure}
\includegraphics[width=0.8\textwidth]{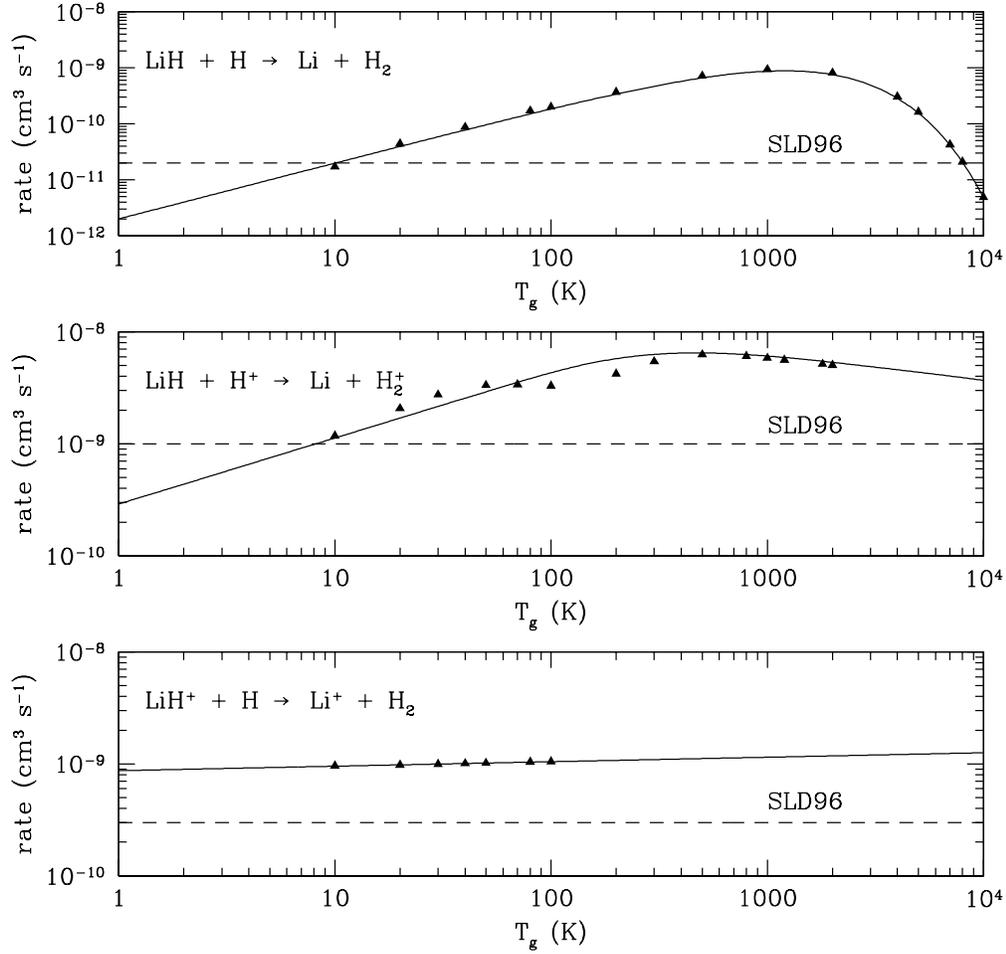}
\caption{A comparison of reaction rates for: destruction of LiH with H 
reaction (top panel) and H$^+$ (middle panel); destruction of LiH$^+$ 
with H (lower panel). The solid lines are fits of the computed data
(open triangles) which used quantum reactive scattering methods (see
Section~2). Dashed curved represent estimates by SLD96.}
\end{figure}
\end{center}

\begin{center}
\begin{figure}
\includegraphics[width=0.8\textwidth]{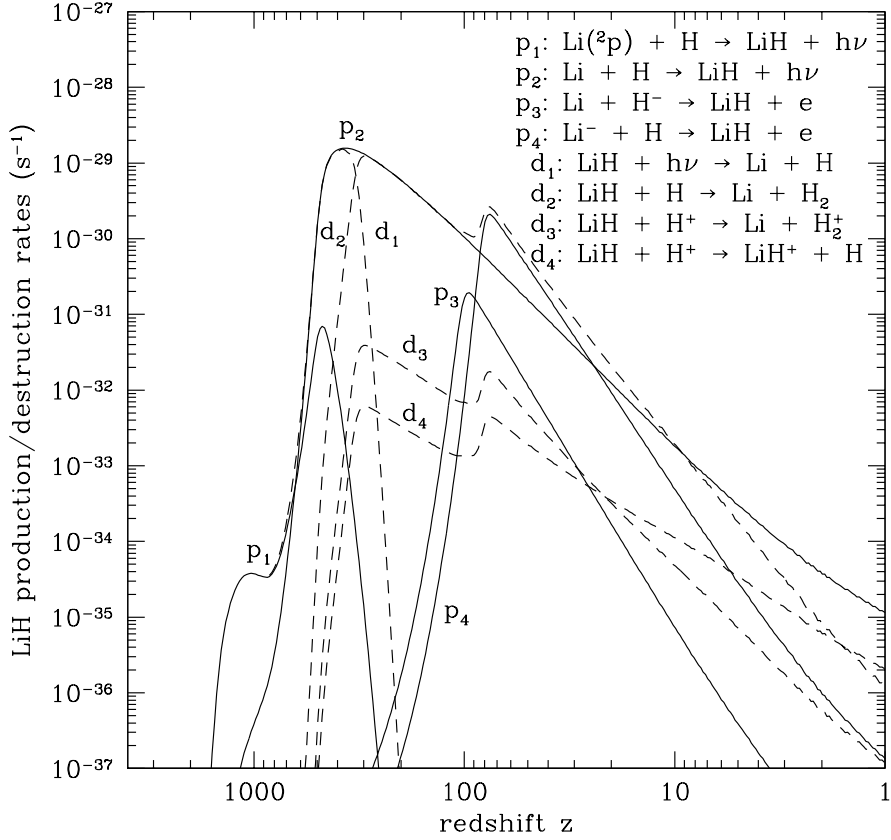}
\caption{Computed production/destruction rates of LiH as function 
of redshift. Solid and dashed curves represent production and 
destruction processes, respectively. See text for details.}
\end{figure}
\end{center}

\begin{center}
\begin{figure}
\includegraphics[width=0.8\textwidth]{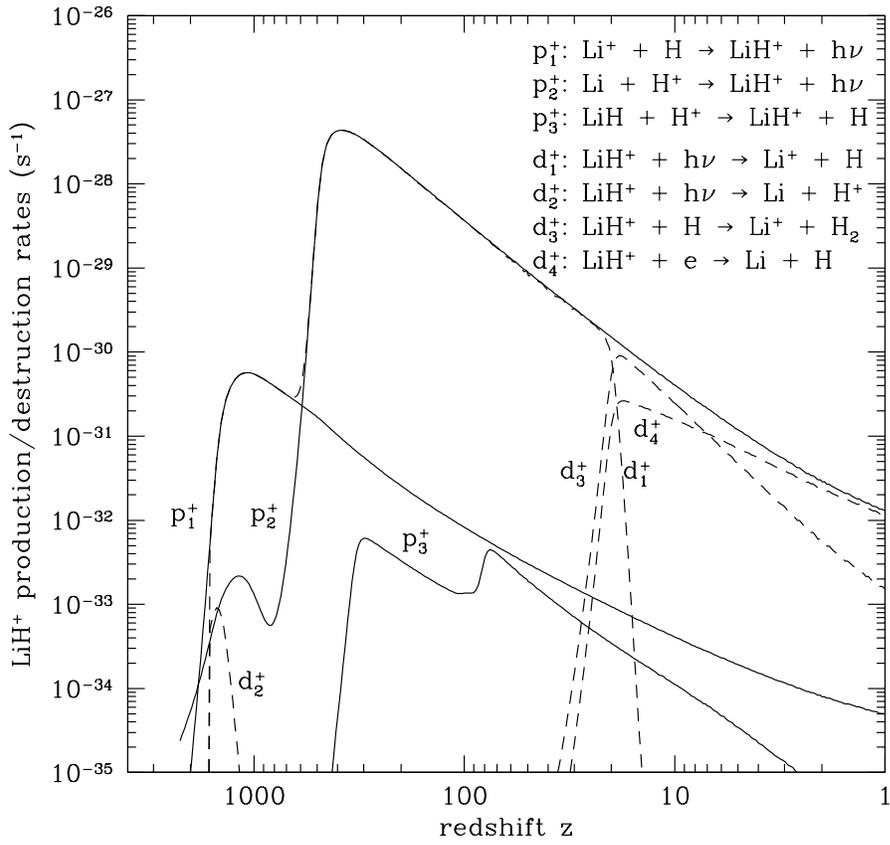}
\caption{Same as Figure~2 for LiH$^+$. See text for details.}
\end{figure}
\end{center}

\begin{center}
\begin{figure}
\includegraphics[width=0.8\textwidth]{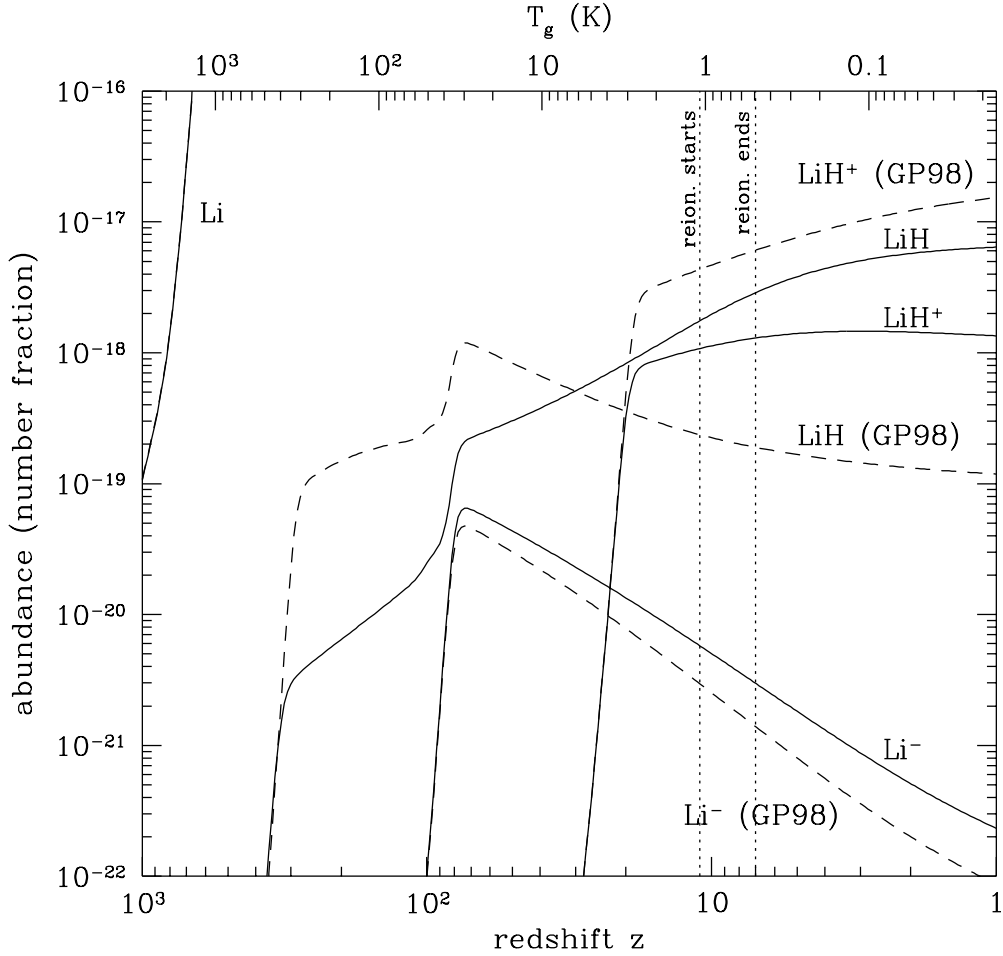}
\caption{Relative abundances of Li, Li$^-$, LiH, and LiH$^+$ in the
post-recombination era as function of redshift $z$ (lower scale) and
gast temperature $T_{\rm g}$ (upper scale): present results (solid curves),
results of GP98 (dashed curves). The dotted lines at $z$=11 and $z=7$ 
indicate the approximate redshifts at which reionization of the primordial
gas starts ($z\simeq 11$) and is completed ($z\simeq 7$), according to 
Spergel et al.~(2007).}
\end{figure}
\end{center}

\end{document}